# Upside Risk Effect on Reliability of Microgrids Considering Demand Response Program and COVID-19: An Investigation on Health System and Power System Interactions

Tohid Khalili, Student Member, IEEE, Seyed Iman Habibi, Student Member, IEEE, Seyyed Ali Ghorashi Khalil Abadi, Student Member, IEEE, Sadaf Ahmadi, Ali Bidram, Senior Member, IEEE

*Abstract*— COVID-19 has a vast impact on the power systems considering the customers' demand and human resources. During this situation, the utilization of microgrids (MGs) may help the power systems balance the generation and consumption of power, which leads to customer satisfaction. In this paper, the optimal power scheduling of energy sources in an islanded MG by considering the upside risk (UR) is proposed for the very first time. The intended islanded MG consists of various sources such as wind turbine (WT), photovoltaic (PV), diesel generator (DGR), and battery. The goals of this work are minimizing the energy not supplied (ENS) in islanded mode considering the COVID-19's effect and implementing the demand response program (DRP). The difference between target ENS and actual ENS when actual ENS is less than the target is defined as UR. The results indicate that the UR related to the ENS of the islanded MG decreases significantly by slightly increasing the ENS. Moreover, COVID-19 decreases the ENS considerably and has a bigger effect than the DRP.

## I. Introduction

Starting in late 2019, millions of individuals were infected or died by COVID-19 according to the World Health Organization (WHO) [1]. COVID-19 has a huge effect on the power system that should be assessed to find its several side effects. In addition, microgrids (MGs) have attracted the attention of researchers and industry in most countries in recent years. Using MGs leads to lower loss and higher reliability of the power system. In recent studies, various sources and structures are used in the MGs. Usually, the sources such as renewable energy sources (RESs) and energy storage systems are used in MGs. On the other hand, the MGs' operators try to have minimum energy not supplied (ENS), regarding the fact that the output power of RESs is stochastic. If the deviation of ENS becomes more than specified values, the MGs' operators will be dissatisfied. This phenomenon is defined as the risk for operators that want to minimize this factor. Several studies are performed in the field of islanded MGs with different aims. In some research projects, different types of applicable designs for the operation of the MGs improvement are reviewed. The control and operation of MGs [2], as well as the energy management methods [3], are the summary of the mentioned studies.

This work is supported by the National Science Foundation EPSCoR Program under Award #OIA-1757207.
Tohid Khalili, Seyed Iman Habibi, Seyyed Ali Ghorashi Khalil Abadi, and Ali Bidram are with the Department of Electrical and Computer Engineering, University of New Mexico, Albuquerque, USA. (e-mails: {khalili, Habibi, ghorashi, bidram}@unm.edu). Sadaf Ahmadi is with the Department of Biology, University of New Mexico, Albuquerque, USA. e-mail: (sadafah54@unm.edu).

In [4], intelligent management of energy storage and optimized operation of MG are raised as optimization problems. Scheduling for an MG consisting of batteries, fuel cells, micro-turbines (MTs), photovoltaics (PVs), and wind turbines (WTs) is investigated in [5]. In [6], energy scheduling of an MG is done in which special attention is paid to the RESs' variable power generation. Operation management of an MG consisting of RESs, which is supported by several sources such as batteries and MTs, is studied in [7].

The studies aiming at the risk concept are investigated in the power system in order to assess the impact of the risk on the various stochastic variables of the MGs. In [8], financial risk minimization of the generation companies (GENCOs) is surveyed during the outage of the different parts of the power system. In addition, the profit sensitivity of GENCOs to the risk, making a balance between GENCOs profit and existing risk is discussed in [9]. Due to uncertainty that exists in the power generation of the RESs, risk in the scheduling and uncertainties in the power trade is considered. In the islanded mode, due to the random and unreliable generation of the RESs and lack of power exchange between MG and the main grid, MG's operator confronts several severe difficulties. In the islanded mode of the MGs, the minimization of ENS is the main objective function of the different papers. MG's optimal power flow of the distributed generators (DGs) for having a stable operation in the islanded mode is proposed in [10].

In this work, MG's optimal power scheduling of RESs and diesel generators (DGRs) is implemented with the goal of risk evaluation. Moreover, minimization of the ENS in the islanded mode is another objective function of this paper. Due to the variations in the generation of the RESs and the load profile, Weibull, Beta, and normal distribution functions are utilized. ENS increment makes MG's operators and the consumers unsatisfied. If the ENS violates the expected determined limit, it causes a phenomenon, called upside risk (UR). Therefore, the aim of this research is ENS minimization in islanded mode and risk evaluation. Also, unit commitment (UC) is performed in order to operate DGRs properly. The simulations are performed for five probable days using GAMS and MATLAB software. In fact, stochastic scheduling is accomplished. The UR is evaluated as well as optimal power scheduling for the various sources is performed. In addition, the effects of the demand response program (DRP) and COVID-19 on the ENS of the islanded MGs are investigated.

The rest of this work is structured as follows: Section II describes the modeling, the problem formulation, and the optimization method. In section III, the results and discussion are presented. Finally, Section IV concludes the paper and declares the main findings.

## II. Problem Formulation

### A. Modeling

In this section, the utilized model and method are presented which are built upon our previous work on this subject [11]. The structure of MG is first introduced in [11], see Fig.1. The considered islanded MG consists of 6 buses. PVs are installed on the first bus and the second bus. WTs are located in the third bus and the DGRs are installed in the fourth, fifth, and sixth buses. In addition, there are batteries in the considered MG. Batteries and load are installed on the sixth bus. In this work, due to the stochastic nature of the load and the power generation of the PVs and WTs, the normal, Beta, and Weibull distribution functions are employed to model their stochastic behavior, respectively [11]. Furthermore, the MG's loads can participate in DRP. Here, the simulations are performed for five sample scenarios (SCs) generated by the various distribution functions attributed to five days.

### B. Problem Formulation

This section consists of two main subsections. The problem constraints are discussed in the first subsection and the objective functions are introduced in the second one.

*1) Constraints*

The considered MG consists of several types of power equipment in which each of them has its own special constraints. These constraints are classified as follows:

- *Equipment*: Constraints of the RESs, DGRs, and batteries derived from [11]
- *UR constraints:* UR is the possibility of asset or value increment beyond the expectations [12]. This concept can show a red flag that a variable is taking many risks. In addition, it can be considered a positive risk that gives more freedom to the operators of the system to manage the system and reach their goals by employing different tools and methods by deciding between various options. UR is related to the MGs operator's willingness to take risks and use the maximum capability of the system to keep the high priority variables between certain margins regarding the risk associated with that decision. Usually, managers who can highly tolerate the risks choose moves with excessive UR, while managers with less tolerance prefer limited UR to keep their normal performance. In addition, having high UR may hurt the MG during unwanted happening such as islanding since the system is solely focused on some predetermined objectives without having enough freedom to react against possible shocks. To the best of the authors' knowledge, the relation between the UR and ENS in an islanded MG is formulated and proposed here for the very first time. MG's operator wants to be able to manage this system through a wide range of decisions in which UR could facilitate this objective. In addition, the islanded MG's operator tries to minimize the total ENS ($TENS_s$) considering a target ($target_s$) for UR ($UR_s$) of each SC ($s$). The definition of this concept is as follows:

$$\text{if } TENS_s \leq target_s, \quad UR_s = target_s - TENS_s \quad (1)$$
$$\text{otherwise}, \quad UR_s = 0.$$

Equation (1) could be redefined as (2) and (3):

$$0 \leq UR_s + (TENS_s - target_s) \leq M \times (1 - W_s) \quad (2)$$

$$0 \leq UR_s \leq M \times W_s \quad (3)$$

where $M$ is a large and positive number. $W_s$ is a binary index for each *SC* and its value is 1 when $TENS_s \leq target_S$.

Based on the proposed definition, the UR for islanded MG aims to minimize the ENS as

$$\sum_{s=1}^{5}(prob_s \times UR_s) \leq \lambda \times EUR \quad (4)$$

In (4), $prob_S$ is the probability of the $s^{th}$ SC, and it is assumed to be the same for all SCs without loss of generality. Also, $\lambda$ is a number between 0 and 1 used for adjusting the risk during the operation. Additionally, $EUR$ is the expected UR of the islanded MG based on the normal operating values.

- *Power balance constraints:* The power balance equation with the aim of ENS investigation is as follows:

$$\sum_{g=1}^{3}P_{g,t,s} + \sum_{i=1}^{6}PV_{i,t,s} + \sum_{j=1}^{2}WT_{j,t,s} + P_{t,s}^{ch,bat} + P_{t,s}^{disch,bat} + ENS_{t,s} = PL_{t,s} \quad (5)$$

In (5), $P_{g,t,s}$, $PV_{i,t,s}$, and $WT_{j,t,s}$ are the output power of the $g^{th}$ DGR, $i^{th}$ PV, and $j^{th}$ WT at the $t^{th}$ hour and $s^{th}$ SC, respectively. Moreover, $P_{t,s}^{ch,bat}$ and $P_{t,s}^{disch,bat}$ are the batteries charged and discharged power at the $t^{th}$ hour and $s^{th}$ SC, respectively. Furthermore, $PL_{t,s}$ and $ENS_{t,s}$ are the demanded load and the ENS of the system at the $t^{th}$ hour and $s^{th}$ SC, respectively.

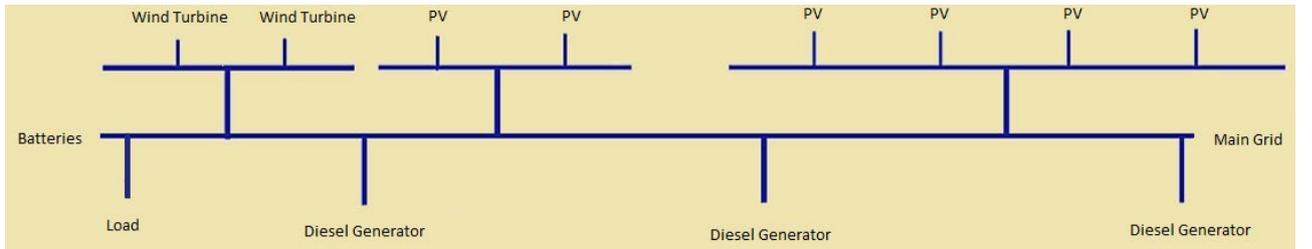

Fig. 1. Studied system

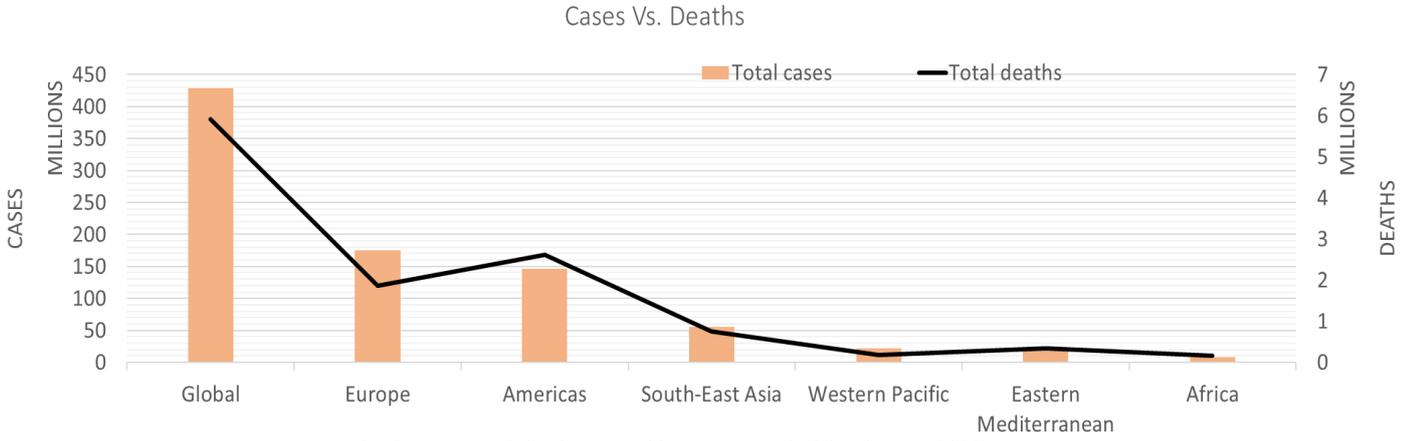

Fig. 2. Cases and deaths caused by COVID-19 (24 February 2022)

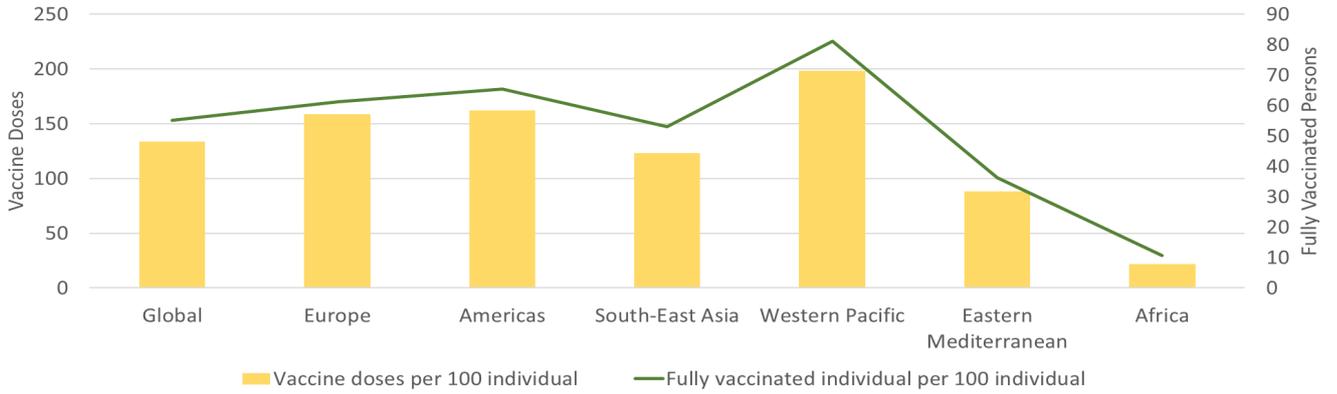

Fig. 3. Vaccination data (21 February 2022)

- *DRP constraints:* Consumers' participation in the management and control of the power system is of particular importance. The islanded MG's operator tends to implement DRP to satisfy its customers and reduce the ENS of the system. From this perspective, the implementation of the DRP improves the reliability of the MGs [13].

  Hence, the DRP is implemented in this study, and its effect on the UR is investigated. In addition, the self-elasticity and the cross elasticity of the load are considered [14]. The main equation of the implemented DRP is provided below and the rest can be found in [11]:

$$-0.15\,PL_{t,s} \leq PL_{t,s}^{D} \leq 0.15\,PL_{t,s} \qquad (6)$$

  In (6), $PL_{t,s}^{D}$ is the load after the DRP implementation at the $t^{th}$ hour and $s^{th}$ SC. It should be mentioned that after the DRP implementation in (5) $PL_{t,s}$ will be substituted with $PL_{t,s}^{D}$.

- *COVID-19 effect:* Millions of people are infected by COVID-19 and inordinate deaths are caused by it. Fig. 2 shows COVID-19 cases and deaths according to WHO [1]. In addition, the number of vaccines administrated in order to prevent COVID-19 spread is indicated in Fig. 3 based on the data provided by WHO [1]. COVID-19 affects the load profile of the system. Based on the U.S. Energy Information Agency (EIA) data and reports the average generation of power is dropped mainly due to COVID-19 [15].

  In this case, a new Coefficient (*CVD*) is defined that is multiplied by the load of the system. *CVD* indicates the average load's change percentage in 2020 compared to 2019. As a result, $PL_{t,s}$ will be substituted with $PL_{t,s}^{CVD}$ in (5). $PL_{t,s}^{CVD}$ is the load affected by COVID-19 and it is defined as follows:

$$PL_{t,s}^{CVD} = (1+CVD) \times PL_{t,s} \qquad (7)$$

*2) Objective Function*

The effect of the UR on the minimization of the ENS is assessed in this work. The following objective function describes the minimization of the *TENS* as below:

$$TENS_s = \sum_{t=1}^{24} ENS_{t,s} \qquad (8)$$

$$TENS = \sum_{s=1}^{5} \left( prob_s \times TENS_s \right) \qquad (9)$$

As a side objective, spilled energy of the system ($SE^{Total}$) including the PVs ($SE_{PV}$) and WTs ($SE_{WT}$) using the maximum output of the PVs ($PV_{i,t,s}^{Max}$) and WTs ($WT_{j,t,s}^{Max}$) are calculated as follows:

$$SE_{PV} = \sum_{s=1}^{5}\left(prob_s \times \sum_{t=1}^{24}\sum_{i=1}^{6}\left(PV_{i,t,s}^{Max} - PV_{i,t,s}\right)\right) \quad (10)$$

$$SE_{WT} = \sum_{s=1}^{5}\left(prob_s \times \sum_{t=1}^{24}\sum_{j=1}^{2}\left(WT_{j,t,s}^{Max} - WT_{j,t,s}\right)\right) \quad (11)$$

$$SE^{Total} = SE_{PV} + SE_{WT} \quad (12)$$

### C. Optimization Method

The optimization method of this work is based on mixed-integer programming (MIP). The GAMS software is used to solve this problem by employing the CPLEX solver. As a result, the obtained results of the simulations are optimal. The proposed method is briefly explained in the following steps:

***Step 1:*** Import the independent initial data.

***Step 2:*** Stochastic production of data (load, RESs maximum output) is performed in MATLAB software and used in the GAMS software.

***Step 3:*** Study the MG for analyzing the ENS value with and without UR. After the production of new random data, the objective function is studied and the MG's ENS for each SC is calculated. Then, the obtained results are compared.

***Step 4:*** Study the MG for surveying the MG's ENS value with and without UR using DRP. The ENS is calculated, and different cross elasticities are considered. Then, the obtained outputs are compared.

***Step 5:*** Study the MG for investigating the MG's ENS value with and without UR implementing the COVID-19 effect. ENS is calculated, and the simulation results are compared.

## III. RESULTS AND DISCUSSION

In this study, ENS minimization of an islanded MG is investigated by applying DRP and COVID-19 effects. In addition to the mentioned considerations, UR is applied to this model to investigate its impact on the ENS of an islanded MG. Furthermore, the spilled energy of the PVs and WTs are also calculated and interpreted. In the considered MG, the maximum output capacity of all PVs installed on the first two buses are 32 kW and 16 kW, respectively. Also, the maximum output capacity of all WTs is 42 kW. Sources capacities are chosen considering the required load of the system [11]. The batteries and DGRs' initial inputs rea derived from [11]. In DGRs, minimum and maximum power generation as well as up rate and down rate of them are considered. Also, UC is implemented on DGRs. In addition, the value of the *CVD* is equal to -2.88% [15].

During the simulations, all outputs and variables of the system including the out of resources as well as batteries charge and discharge schedule are computed, but only the main outputs of the simulation are reflected in the simulation results section to highlight the main contributions of the presented work.

Here, the ENS of the islanded MG is calculated without activating UR's effect on the system and the results are presented in Table I. Also, the target ENS of this case is shown. Moreover, the values of the UR without activation of its related constraints are just calculated for comparison purposes that are called passive UR values.

TABLE I. ENS, Target, and passive UR values [kWh]

| SC | ENS | Target | Passive UR |
|---|---|---|---|
| SC1 | 0 | 9 | 9 |
| SC2 | 8.43 | 9 | 0.57 |
| SC3 | 11.887 | 9 | 0 |
| SC4 | 11.78 | 9 | 0 |
| SC5 | 5.535 | 9 | 3.465 |
| Average | 7.5265 | 9 | 2.607 |

Table I indicates that each SC has its own specific output and the variety of the input resulted in different outputs. As it can be seen, the islanded MG has ENS in some of the SCs which is an undesired event. In addition, the passive UR values present that when the system has zero ENS, the passive UR's value is the maximum which indicates this positive risk maximum amount. On the other hand, when the system has more ENS than the target value, the passive UR values are zero making the minimum value of this positive risk concept. In the next step, UR constraints are activated and different λ values are utilized to get a comprehensive perspective of the proposed model. The results of the ENS minimization accompanied by the UR are illustrated in Fig. 4.

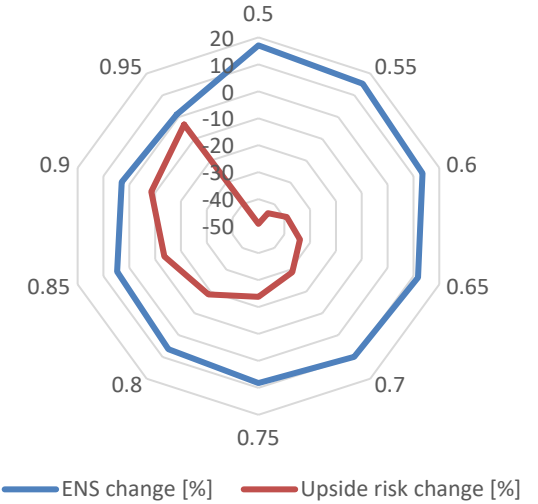

Fig. 4. Change in the ENS and UR values with λ variation

As shown in Fig. 4, by a step-by-step increase of λ from 0.5 to 0.95, the average ENS percentage change compared to the base case in Table I decreases from 17.03% to 1.19%. Moreover, the average UR decreases by employing the UR constraint in (4). The average UR is around -49.18% and -3.343% less than the base case in Table I for 0.5 and 0.95 of λ, respectively. As seen, UR And ENS act against each other i.e., an increase in one of them causes a decrease in the other one. Next, the DRP is applied with passive UR. The results of this application are demonstrated in Table II.

TABLE II. ENS, Target, and passive UR values with DRP [kWh]

| SC | ENS | Target | Passive UR |
|---|---|---|---|
| SC1 | 0 | 5 | 5 |
| SC2 | 4.101 | 5 | 0.899 |
| SC3 | 2.504 | 5 | 2.496 |
| SC4 | 10.925 | 5 | 0 |
| SC5 | 1.964 | 5 | 3.036 |
| Average | 3.8988 | 5 | 2.2862 |

Table II shows that implementation of the DRP reduces the ENS of the islanded MG notably. The average ENS of the system is reduced by about 48.2% compared to Table I. Per the current situation of the system, a new target for the ENS and UR is considered in order to fulfill the real situation of the system. The passive URs are also indicated to be compared with the actual URs in the next parts. After finding the ENS and passive UR values, the UR constraints are activated, and their related results are compared with the passive URs and the previous ENS demonstrated in Table II.

Regarding Fig. 5, the ENS increment, and UR reduction follow Fig. 4's trend as decreasing in ENS reduces the UR. It should be mentioned that the reduction in the UR values is far more than the decrement of the ENS. In this part of the simulations, COVID-19's effect on the islanded MG is examined and the results are reported.

TABLE III. ENS, Target, and passive UR values with DRP [kWh]

| SC | ENS | Target | Passive UR |
|---|---|---|---|
| SC1 | 0 | 4 | 4 |
| SC2 | 3.669 | 4 | 0.331 |
| SC3 | 6.012 | 4 | 0 |
| SC4 | 7.675 | 4 | 0 |
| SC5 | 0 | 4 | 4 |
| Average | 3.4712 | 4 | 1.6662 |

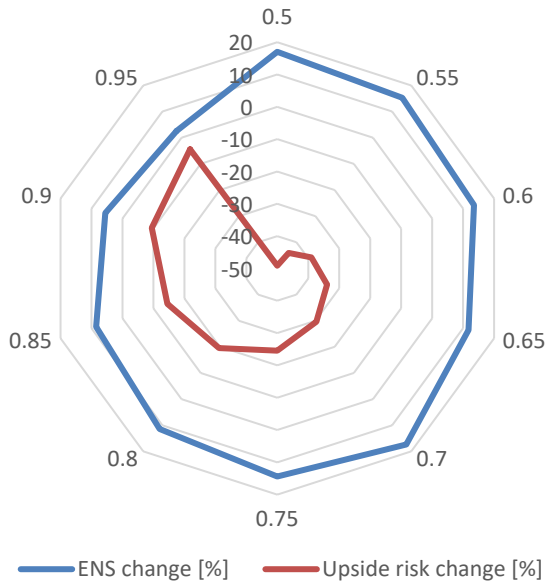

Fig. 5. Change in the ENS and UR values with λ variation with DRP

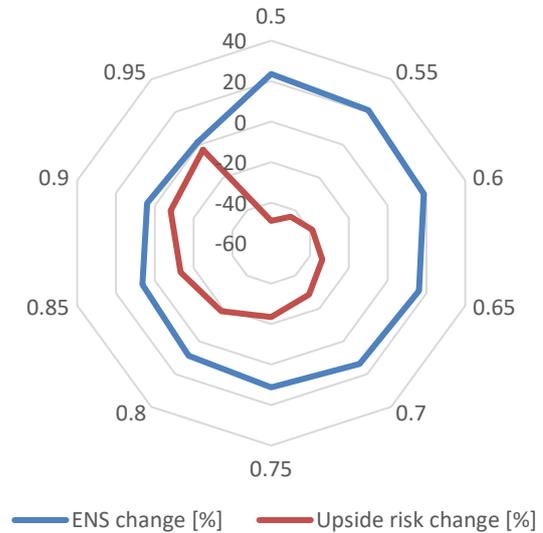

Fig. 6. Change in the ENS and UR values with λ variation with COVID-19 effect

Table III presents the obtained results according to COVID-19's implementation. Table III shows COVID-19's effect on the ENS and passive UR values. As seen, the ENS of the system dramatically decreased by about 53.88% compared to Table II. Since COVID-19 and its related events caused a noticeable reduction in the load profile, the ENS of the system is reduced accordingly. Like Table II, a new target is considered regarding the amended condition of the system. Moreover, the passive UR values are indicated in Table III as well.

In the following part, the UR constraints are activated, and the obtained simulation results are compared with Table III results illustrated in Fig. 6.

Fig. 6 shows a higher change in the ENS values compared to Fig. 5 and the trend of the λ effect stayed the same. As it can be seen, COVID-19 could have a considerable effect on the ENS and UR even in some cases more effective than the demand management methods.

Spilled energy is the amount of energy that the system could use but waste with the incapability of finding a way to exploit it. That said, the RESs' spilled energy of the studied islanded MG are also investigated, and the results are presented in Table IV. This Table includes the with and with UR cases as well as with and without UR cases accompanied by the DRP. Additionally, spilled energy with implementing the COVID-19's effect with and without UR is also presented and compared. As Table IV shows, the majority of the spilled energy in all cases is due to the PVs. Therefore, WTs form a smaller part of the $SE^{Total}$. This could be mainly because of the short generation time of the PVs during the day. In order to investigate the effect of the UR, 0.95 is selected as the value of λ for comparison purposes with other cases.

TABLE IV. Spilled energy in different cases [kWh]

| # | Without UR | UR (λ=0.95) | DRP and without UR | DRP and with UR (λ=0.95) | COVID-19 and without UR | COVID-19 and with UR (λ=0.95) |
|---|---|---|---|---|---|---|
| $SE_{PV}$ | 594.867 | 554.1 | 562.802 | 542.35 | 629.064 | 592.53 |
| $SE_{WT}$ | 159.389 | 216.16 | 123.225 | 189.00 | 150.083 | 258.35 |
| $SE^{Total}$ | 754.256 | 770.26 | 686.027 | 731.35 | 779.147 | 850.89 |

According to this table, UR increases the $SE^{Total}$ in all cases with activated UR constraints. However, UR reduced the $SE_{PV}$ and increased the $SE_{WT}$, particularly. This indicates that UR helped the system move toward a fair spilled energy status between the RESs. On the contrary, DRP significantly reduced the spilled energy of both PVs and WTs. As a result, the $SE^{Total}$ is reduced by about 9.05% and 5.05% without and with UR cases, respectively. On the other hand, COVID-19's effect has a negative impact on the spilled energy, and it increased the $SE^{Total}$ by about 3.3% and 10.47% without and with the UR, respectively. Another point is that the portion of the PVs in the $SE^{Total}$ is increased by applying the COVID-19's effect against the WTs portion.

IV. CONCLUSION

In this paper, the optimal power scheduling for the sources in an islanded MG is proposed concerning the ENS minimization. The main aims of this study are ENS minimization and UR evaluation in an islanded MG. The secondary goals are the DRP implementation and analyzing the effect of COVID-19. All of the simulations are implemented in the GAMS and MATLAB software and the UC is also considered on the DGRs. Different studies are conducted in the paper. First, the studies are performed with and without considering the UR. Then, the DRP is added to the optimization problem with and without UR. Finally, COVID-19's impact on the islanded MG's performance is analyzed. Comparisons show that DRP can remarkably decrease the ENS on average by about 48.2%. Implementation of COVID-19 caused an even higher reduction in ENS values. The case with COVID-19's effect can reach an extraordinary number according to the ENS reduction (on average about 53.88%).

In addition, the simulation results prove that UR and ENS are against each other. In other words, when positive risk decreases the ENS increases and vice versa. Moreover, the spilled energy of the PVs and WTs in different simulation cases is analyzed. The results show that UR could increase the spilled energy, DRP could decrease the spilled energy, and COVID-19 could increase the spilled energy depending on the system's characteristics and the operator's management policies. Using this study, MG operators may use this new method in order to manage/amend their policies to gain more benefits while operating the system appropriately.